\documentclass[aps,pra,reprint,twocolumn,showpacs,floatfix,superscriptaddress]{revtex4-1}

\usepackage{amssymb,amsmath,amstext}                
\usepackage{graphicx}                                               
\usepackage{epstopdf}                                               
\usepackage{color}                                                     
\usepackage{bm}                                                        
\usepackage{appendix}                                              
\usepackage[utf8]{inputenc}
\usepackage{bbold}
\usepackage{bbm}
\usepackage{latexsym}
\usepackage[T1]{fontenc}
\usepackage{xcolor}
\usepackage{braket}
\definecolor{lblue} {RGB}{51,71,158}
\usepackage[colorlinks=true,citecolor=blue,linkcolor=blue,urlcolor=lblue]{hyperref}

\usepackage[shortlabels]{enumitem}

\usepackage[normalem]{ulem}

\definecolor{darkgreen}{rgb}{0.13, 0.55, 0.13}

\def\beq{\begin{equation}}
\def\eeq{\end{equation}}

\usepackage{dcolumn}
\usepackage{sidecap}
\usepackage[caption=false]{subfig}
\usepackage{bm}

\begin{document}

\title{Finite-Size scaling analysis of many-body localization transition \\ in quasi-periodic spin chains}

\author{Adith Sai Aramthottil}
\email{adithsai.a@doctoral.uj.edu.pl}
\affiliation{Institute of Theoretical Physics, Jagiellonian University in Krak\'ow, \L{}ojasiewicza 11, 30-348 Krak\'ow, Poland}
\author{Titas Chanda}
\affiliation{Institute of Theoretical Physics, Jagiellonian University in Krak\'ow, \L{}ojasiewicza 11, 30-348 Krak\'ow, Poland}
\affiliation{The Abdus Salam International Center for Theoretical Physics (ICTP), Strada Costiera 11, 34151 Trieste, Italy}
\author{Piotr Sierant}
\affiliation{The Abdus Salam International Center for Theoretical Physics (ICTP), Strada Costiera 11, 34151 Trieste, Italy}
\affiliation{ICFO-Institut de Ci\`encies Fot\`oniques, The Barcelona Institute of Science and Technology, Av. Carl Friedrich
Gauss 3, 08860 Castelldefels (Barcelona), Spain}
\author{Jakub Zakrzewski}
\email{jakub.zakrzewski@uj.edu.pl}
\affiliation{Institute of Theoretical Physics, Jagiellonian University in Krak\'ow, \L{}ojasiewicza 11, 30-348 Krak\'ow, Poland}
\affiliation{Mark Kac Complex Systems Research Center, Jagiellonian University in Krak\'ow, \L{}ojasiewicza 11, 30-348 Krak\'ow, Poland}

\date{\today}

\begin{abstract}
We analyze the finite-size scaling of the average gap-ratio and the entanglement entropy across the many-body localization (MBL) transition in one dimensional Heisenberg spin-chain with quasi-periodic (QP) potential. 
By using the recently introduced cost-function approach, we compare different scenarios for the transition  using exact diagonalization of systems up to 22 lattice sites.
Our findings suggest that the MBL transition in the QP Heisenberg chain belongs to the class of Berezinskii-Kosterlitz-Thouless (BKT) transition, the same as in the case of uniformly disordered systems as advocated in recent studies. 
Moreover, we observe that the critical disorder strength shows a clear sub-linear drift with the system-size as compared to the linear drift seen in random disordered models, suggesting that the finite-size effects in the MBL transition for the QP systems are less severe than that in the random disordered scenario.
Moreover, deep in the ergodic regime, we find an unexpected double-peak structure of distribution of on-site magnetizations that can be traced back to the strong correlations present in the QP potential.
\end{abstract}

\maketitle


\section{Introduction}
\label{sec:intro}

Generic isolated quantum many-body systems are expected, according to {the} eigenstate thermalization hypothesis \cite{Deutsch91,Srednicki94,Rigol08}, to approach equilibrium described by an appropriate statistical ensemble determined by a few global integrals of motion such as energy or total momentum -- see~\cite{Alessio16, Vidmar16}. One exception to this hypothesis is provided by the phenomenon of many-body localization (\textbf{MBL}) \cite{Basko06,Gornyi05} which occurs in presence of strong disorder and interactions. In this dynamical phase the approach to equilibrium is inhibited and the system indefinitely preserves detailed information about its initial state  due to presence of local integrals of motion \cite{Serbyn13b,Huse14,Ros15,Mierzejewski18}. In consequence, the transport is slowed down and eventually 
suppressed \cite{Nandkishore15, Znidaric16, Alet18, Abanin19}, and the entanglement spreads slowly \cite{Znidaric08,Serbyn13a,iemini2016signatures}.

The combination of strong disorder and interactions makes the phenomenon of MBL tractable only in exact numerical calculations for relatively small lattice systems \cite{Pietracaprina18, Sierant20p}. This makes understanding of ergodic to MBL transition  a formidable undertaking, especially in conjunction with non-perturbative mechanisms of delocalization of MBL phase \cite{DeRoeck16, DeRoeck17,DeRoeck17b}. Despite intensive research \cite{Oganesyan07, Pal10,Bera15, Luitz15, Mondaini15,Khemani17a, Enss17, Bera17, Doggen18, Chanda20t, Herviou19, Colmenarez19, Sierant20p, Vidmar21} the status of the MBL phase is not fully understood as shown by the recent debate about {the} stability of MBL \cite{Suntajs20e, Sierant20b, Abanin19a, Panda19, Kiefer20, Luitz20, Sels-P21, Morningstar21, Sels21}.
On the other hand, the existence of MBL is essentially certified in sufficiently strongly disordered spin chains as shown in \cite{Imbrie16,Imbrie16a}.

Two primary hypotheses proposed to describe the MBL transition assume that the correlation length $\xi$ in the system:
\begin{enumerate}[(A)]
 \item diverges at the transition in a power-law fashion  $\xi_0 = \frac{1}{\vert W-W^*\vert^{\nu}}$, where $\nu$ is a critical exponent and $W^*$ is the critical disorder strength \label{eq:power_law}
 \item assumes a  Berezinskii–Kosterlitz–Thouless  ({\textbf{BKT}}) scaling: $\xi_{BKT} = \exp\biggl\lbrace\frac{b_{\pm}}{\sqrt{\vert W-W^*\vert}}\biggl\rbrace$, where $b_{\pm}$ are non-universal parameters on the two sides of the transition. \label{eq:BKT}
\end{enumerate}
The hypothesis \ref{eq:power_law} of power-law divergence of $\xi$ was supported by the early real-space renormalization group approaches \cite{Vosk15, Potter15}.
However, the majority of numerical studies typically  {found} the critical exponent $\nu \sim 1$  \cite{Kjall14, Luitz15} violating the Harris bound $\nu > 2$ \cite{Harris74,Chayes86, Chandran15a}. An exception is provided by the system size scaling of the Schmidt gap \cite{Gray18}.
The hypothesis of BKT scaling \ref{eq:BKT} is supported by real-space renormalization group approaches \cite{Goremykina19,Dumitrescu19,Morningstar19,Morningstar20} based on the avalanche scenario of delocalization of MBL phase \cite{DeRoeck17, Luitz17} and advocated by the recent numerical studies \cite{Suntajs20,Laflorencie20, Hopjan21}.

The phenomenon of MBL may also occur when the random disorder (\textbf{RD}) in the system is replaced by a quasi-periodic (\textbf{QP}) potential \cite{Iyer13, Naldesi16, Setiawan17,BarLev17, Bera17a,Weidinger18, Doggen19, Weiner19, Mace19}. MBL in QP systems was studied in a number of experimental settings \cite{Schreiber15,Luschen17, Rispoli19,Leonard20} 
(we note that the RD could, in principle, be introduced in such systems via a speckle potential \cite{Maksymov20}). QP potential has a period incommensurate with the lattice constant, hence it breaks the translational invariance effectively acting as a disorder. However, the strong long-range correlations present in the QP potential may severely affect the properties of the MBL transition \cite{Khemani17,Zhang18}, 
{and lead to significantly smaller variations in system properties from one disorder realization to another one as  compared to the RD scenario}
\cite{Khemani17,Sierant19b}. 
Furthermore, due to lack of local fluctuations in the QP potential, a mechanism giving rise to the ergodic seeds that could initialize the avalanches delocalizing the MBL phase remains to be identified \cite{Gopalakrishnan20}.
So far, studies of MBL in QP systems concentrated mainly on the scenario  \ref{eq:power_law} for the MBL transition, finding a critical exponent $\nu \sim 1$ \cite{Khemani17,Lee17,Agrawal20} apparently satisfying the Harris-Luck criterion ($\nu > 1$) \cite{Luck93}. This value of the critical exponent was not confirmed by the real-space renormalization group calculation of \cite{Zhang18} which finds $\nu \sim 2.4$. Moreover, the recent examination of local integrals of motion in the QP systems \cite{Singh21} suggest that $\nu \gtrsim 2$ and that the critical disorder strength for the transition to MBL is much larger than previously expected.

Motivated by these results we decided to perform a quantitative comparison of the scalings \ref{eq:power_law} and \ref{eq:BKT} in QP systems. 
To that end we follow the cost-function approach proposed in \cite{Suntajs20} which allows also for a direct comparison with the RD case.
Contrary to the expectations \cite{Gopalakrishnan20}, we find that the BKT scaling leads to better finite-size collapses of the data for QP model than the hypothesis of power-law divergence of correlation length. In either of the cases we find that the drift of the critical disorder strength with system size $L$ is weaker than in systems with RD and slows down with increase of $L$. Moreover, we find an unexpected behavior of spin correlation functions in the ergodic phase of the model that we trace back to strong correlations in the QP potential.

The rest of this work is structured as follows. {In Sec. \ref{sec:model}, we give brief introduction to the QP system considered in this work and to the corresponding physical quantities used for the analysis of the ergodic-MBL transition. We discuss the appearance of an exotic double-peak structure in the distribution of on-site spin expectation values deep in the ergodic phase in Sec.~\ref{sec:double-peak}.
Section~\ref{sec:scaling} is devoted for the core of our results, where we present the detailed analysis of the finite-size scaling across the MBL transition in the QP system. Finally, we draw our conclusions in Sec.~\ref{sec:conclu}.
}

\section{Model and Observables}
\label{sec:model}

We consider the paradigmatic  model for MBL studies, namely the 1D Heisenberg spin-$1/2$ chain of length $L$ with the following Hamiltonian
\begin{equation}
\hat{H} = \sum_{l=1}^{L-1}\left[ \hat{S}_l^x \hat{S}_{l+1}^x + \hat{S}_l^y \hat{S}_{l+1}^y + \hat{S}_l^z \hat{S}_{l+1}^z \right] + \sum_{l=1}^L h_l \hat{S}_l^z,
\label{eq:ham}
\end{equation}
where $\hat{S}_l^{\alpha}$ ($\alpha = x, y, z$) are the spin-$1/2$ operators and $h_l$ denote  the on-site potentials. In this 
work, we consider the QP Heisenberg chain where the on-site potentials $h_l$ are given as
\begin{equation}
\label{qppot}
h_l = (W/2) \cos(2\pi k l + \phi).
\end{equation}
We fix $k$ as {the inverse} golden ratio $k=(\sqrt{5}-1)/2$ and $\phi$ is a random phase taken from the uniform distribution between  $[0,2\pi)$. 
The amplitude $W$ of the QP field
plays a role of the `disorder strength' in the considered system. Hence, we sometimes refer to the QP potential \eqref{qppot} for a given value of $\phi$ as a `disorder realization', and refer to the average over $\phi$ as the `disorder average'.
We note that the QP disorder calls for the enforcement of open boundary condition.

To investigate a crossover between the ergodic and the MBL regimes, we consider two widely used observables, namely the half-chain entanglement entropy (\textbf{EE}) $\mathcal{S}$ of the eigenstates \cite{Luitz15, Yu16} and the average gap-ratio $\bar{r}$ \cite{Oganesyan07, Atas13}.
For a given disorder realization, the energy-level gap-ratios are defined as 
\begin{equation}
r_n \equiv \frac{\min\lbrace\Delta_n,\Delta_{n+1}\rbrace}{\max\lbrace\Delta_n,\Delta_{n+1}\rbrace},
\end{equation} 
where $\Delta_n = E_n-E_{n+1}$ are the spacings between subsequent eigenenergies. The average gap-ratio $\bar{r}$  
is then obtained by first averaging $r_n$ within a given disorder realization and, subsequently, over different disorder realizations.
The average gap-ratio is equal to the random matrix theory (\textbf{RMT}) prediction $\bar{r}\approx 0.531$ in the fully delocalized regime of  models preserving the generalized time reversal symmetry (the case we consider here), while for fully localized systems it takes the value characteristic for Poisson distribution $\bar{r} \approx0.386$ \cite{Atas13}. 

The half-chain EE is defined as the von Neumann entropy of the reduced density matrix $\rho_{L/2}$ as follows
\begin{equation}
\mathcal{S}_{L/2} = -\text{Tr}[\rho_{L/2} \ln \rho_{L/2}],
\end{equation}
where $\rho_{L/2} = \text{Tr}_{1, 2, ..., L/2} \ket{\psi} \bra{\psi}$ is obtained by tracing out half of the system.
To minimize the system-size dependence of the EE, 
we rescale it by the corresponding  RMT value $\mathcal{S}_{RMT}=(L/2)\ln(2)+(1/2+\ln(1/2))/2-1/2$ as $\mathcal S=\mathcal{S}_{L/2}/\mathcal{S}_{RMT}$ \cite{Vidmar17, Huang19b, Huang21}. Finally, as in the case for $\bar{r}$, we average the rescaled EE over eigenstates corresponding to a particular disorder realization, and then over different disorder realizations.

\begin{figure}%
\includegraphics[width=\linewidth]{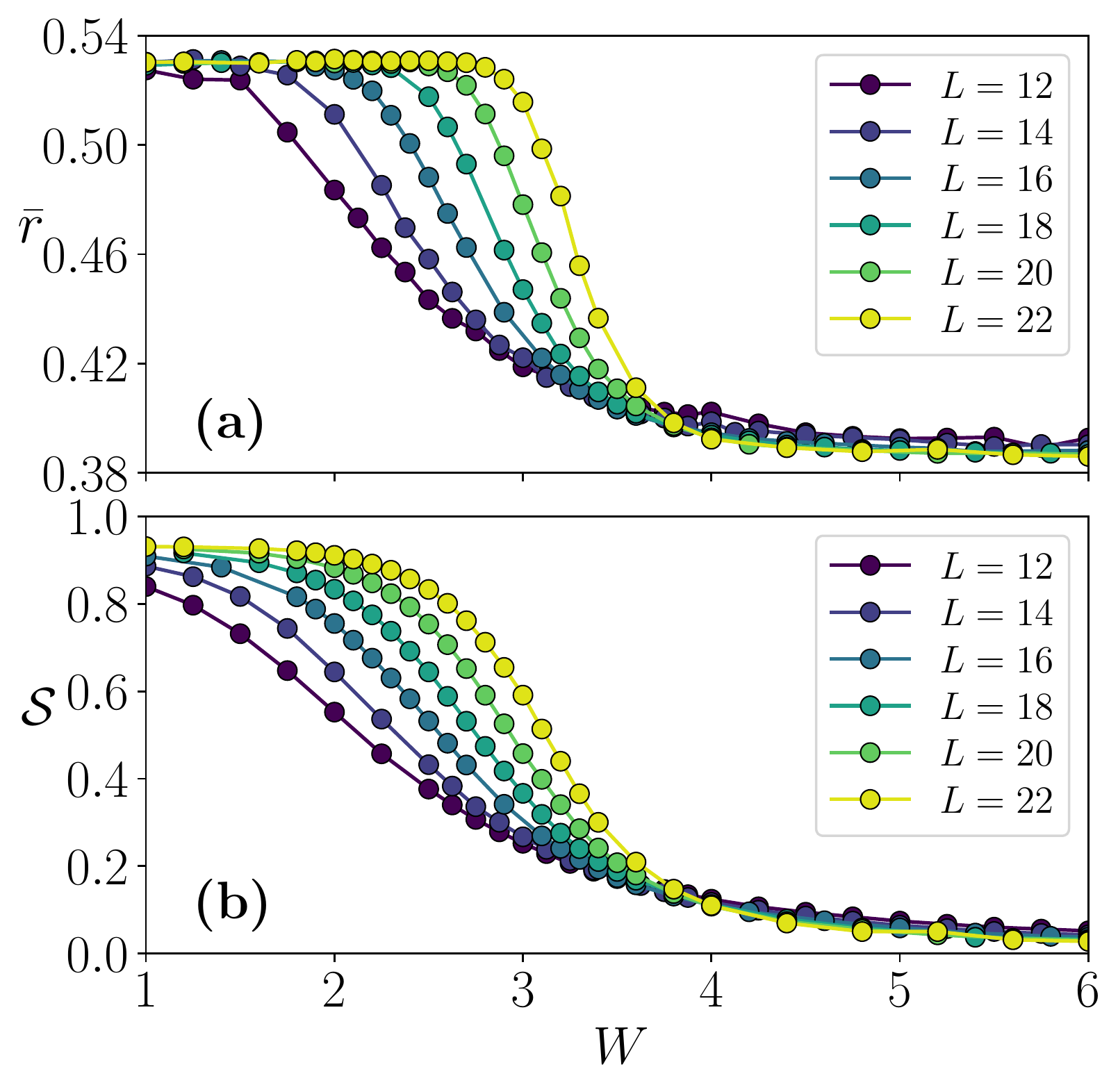}
\caption{The dependence of (a) the average gap-ratio $\bar{r}$ and (b) the half-chain EE $\mathcal{S}$ on the amplitude $W$ of the QP disorder 
for different system sizes $L \in [12, 22]$.}
\label{fig:basicplots}
\end{figure}

 \begin{table}[htb]
 	\caption{\label{tab:table3}%
 The number of eigenstates and the number of realizations used in this work.}
 	\begin{ruledtabular}
 		\begin{tabular}{lcdr}
 			\textrm{$L$}&
 			\textrm{No. of eigenstates}&
 			\multicolumn{2}{c}{\textrm{No. of realizations}}\\
 			\colrule
 			10 & 252 & \multicolumn{2}{c}{30000} \\
 			12 & 300 & \multicolumn{2}{c}{2000} \\
 			14 & 340 & \multicolumn{2}{c}{5000} \\
 			16 & 720 & \multicolumn{2}{c}{2000} \\
 			18 & 900 & \multicolumn{2}{c}{2000} \\
 			20 & 1000 & \multicolumn{2}{c}{1000} \\
 			22 & 1000 & \multicolumn{2}{c}{1000} \\
 		\end{tabular}
 	\end{ruledtabular}
 \end{table}

In this work, we consider the system \eqref{eq:ham} of  size $L \in [10, 22]$, and vary the disorder strength $W$ within the range $[0.4, 6]$. 
The variations of the average gap-ratio and the half-chain EE with increasing  $W$, as the system undergoes a crossover between the ergodic  to the MBL regimes, are shown in Fig.~\ref{fig:basicplots} for $L \in [12, 22]$.  
At each particular $W$ and system size $L$, the observables are calculated by averaging over eigenstates near 
the rescaled energy
 \begin{align}
 \epsilon = \frac{E-E_{\min}}{E_{\max}-E_{\min}}= 0.5
 \end{align}
for a given disorder realization, and subsequently over different realizations demarcated by different random phases $\phi$.
The numbers of eigenvalues and disorder realizations used in this study
for each length $L$ are enumerated in Table \ref{tab:table3}.
For system size $L \leq 14$ we use standard exact diagonalization (ED) method for dense matrices, 
while for $L \geq 16$ recently developed \textit{polynomially filtered exact diagonalization} (\textbf{POLFED}) method \cite{Sierant20p} is used.
{The results of both algorithms were compared between each other for $L=16$ and also verified with the standard ``shift-and-invert'' diagonalization scheme \cite{Pietracaprina18}. No difference up to machine precision has been detected.}

\section{Non-Gaussian behavior of $\braket{\hat{S}^z_l}$ in the ergodic regime}
\label{sec:double-peak}

\begin{figure}%
\includegraphics[width=\linewidth]{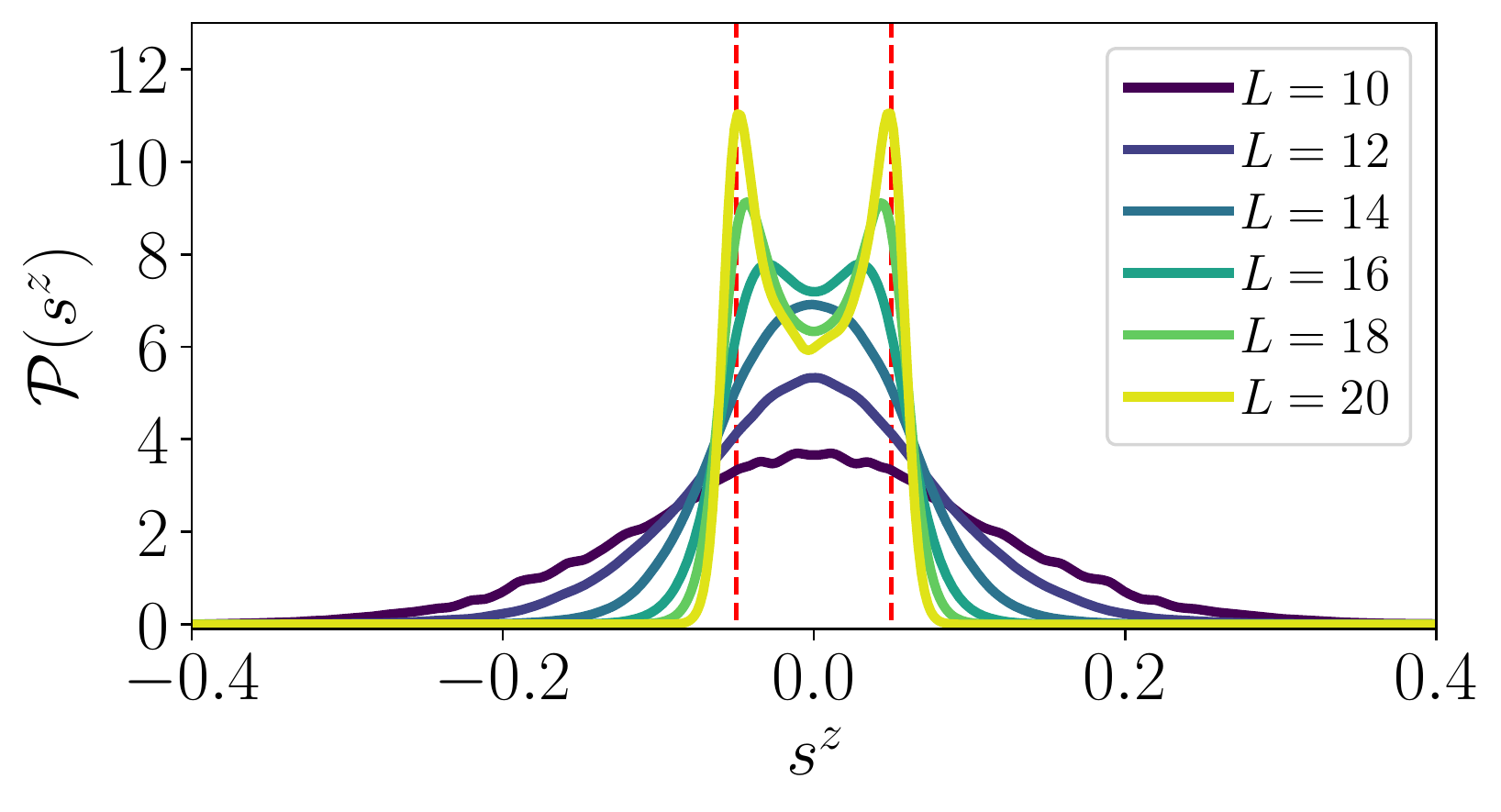}
	\caption{The probability distribution $\mathcal{P}(s^z)$ for site-resolved spin expectation values $s^z$
 with field strength $W=1$. We find for $L\geq 16$ double peaks start to appear. The red dashed lines are drawn at $s_z= \pm 0.05$ to give an estimate of the location of peaks for $L=20$.}
	\label{fig:mag}
\end{figure}

In the QP system considered here, the site dependent potential $h_l$ is fully correlated and its values on all lattice sites are determined by the value of the phase $\phi$.  In this section we demonstrate that this property of the QP potential affects the system properties profoundly even  deep {in} the ergodic regime.

For the standard Heisenberg model with RD, the probability distribution $\mathcal{P}(s^z)$ for site-resolved spin expectation values $s^z_l = \braket{\hat{S}^z_l}$ for mid-energy eigenstates ($\epsilon \approx 0.5$) has a U-shape with peaks at $s^z \simeq \pm 0.5$ in the MBL phase due to the presence of local integrals of motion that have substantial overlaps with $\hat{S}^z_l$ operators \cite{Khemani16,Lim16, Dupont19,Hopjan19,Laflorencie20}.
According to {the} eigenstate thermalization hypothesis, $\mathcal{P}(s^z)$ has a Gaussian distribution with a peak at $s^z = 0$ and variance that quickly decreases with $L$ deep in the ergodic regime. In the intermediate regime between ergodic and MBL phases, the distribution of $s^z$ acquires long tails, with bulk of the distribution concentrated in the peak at $s^z = 0$. Both types of behavior were observed for Heisenberg model in e.g., \cite{ Luitz16b, Luitz16c,Colmenarez19, Laflorencie20}.

Fig.~\ref{fig:mag} shows the probability distribution $\mathcal{P}(s^z)$ of $s^z$ for system sizes $L \in [10, 20]$ and disorder strength $W=1$ that lies deep inside the ergodic regime (for $W=1$ the average gap ratio is $\overline r \approx 0.531$ and $\mathcal S \approx 1$, see Fig.~\ref{fig:basicplots}). 
Strikingly, for larger system sizes, $\mathcal{P}(s^z)$ is not Gaussian anymore. Rather,  $\mathcal{P}(s^z)$ develops characteristic peaks that appear at non-zero $s^z$ (e.g., $s^z \simeq \pm 0.05$ for $L=20$).

\begin{figure}%
\includegraphics[width=\linewidth]{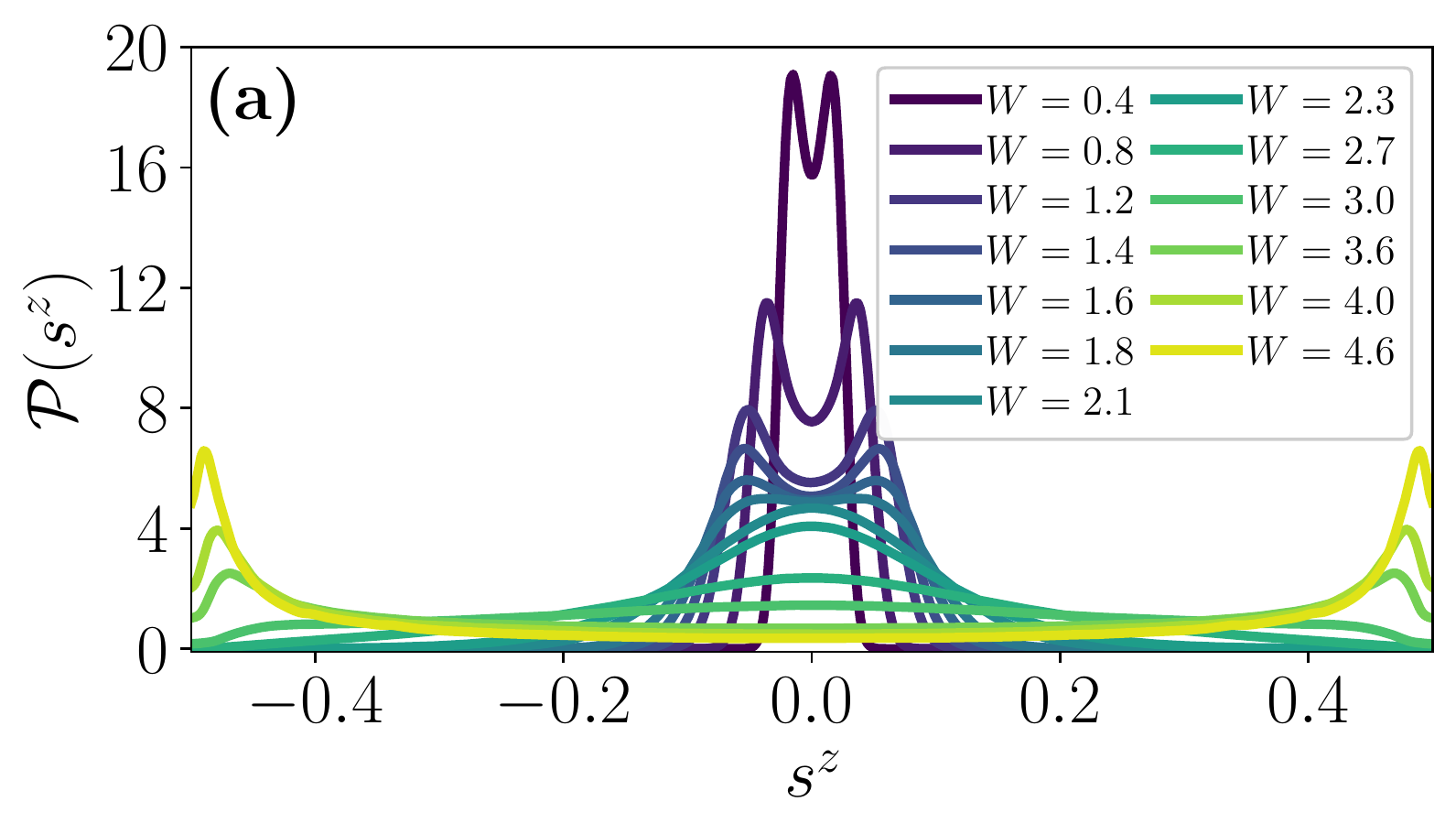}
\includegraphics[width=\linewidth]{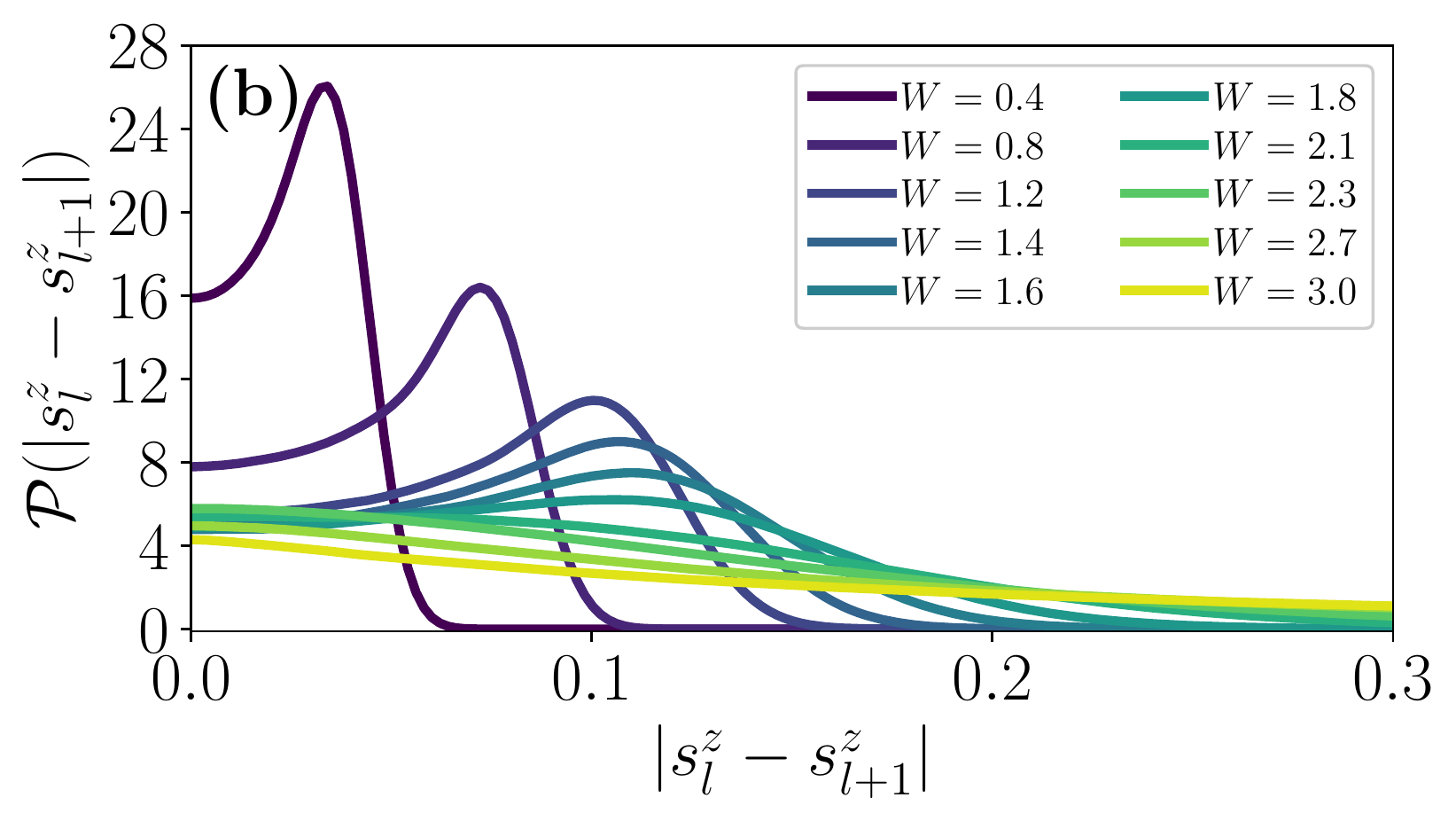}
	\caption{(a) The probability distribution $\mathcal{P}(s^z)$ for site-resolved spin expectation values $s^z$
with varying field strength $W$ for the system-size $L=18$. For this system-size, the characteristic double-peak nature of the probability distribution disappears at $W \simeq 2$ and the U-shaped pattern corresponding to the MBL phase starts to appear for $W \gtrsim 3.5$.
(b) The probability distribution $\mathcal{P}(|s^z_{l} - s^z_{l+1}|)$ for the nearest-neighbor spin-differences $|s^z_{l} - s^z_{l+1}|$ the with varying field strength $W$ for the system-size $L=18$. Interestingly, $\mathcal{P}(|s^z_{l} - s^z_{l+1}|)$ attains its maximum value at non-zero $|s^z_{l} - s^z_{l+1}|$ deep in the ergodic regime for $W \lesssim 2$ for this particular system-size.}
\label{fig:mag_hist_2}
\end{figure}

On closer inspection, the double-peak structure in $\mathcal{P}(s^z)$ appears deep in the ergodic regime for sufficiently large system size -- it is apparent already for $W=0.4$ for $L=18$, as shown in Fig.~\ref{fig:mag_hist_2}(a). With increase of the QP potential amplitude $W$, the distance between peaks increases. At the same time, the central minimum gets shallower and, upon further increase of $W$, the double peak structure  slowly vanishes while the system is still in the ergodic regime. For example, for $L=18$, the double-peak disappears at around $W \simeq 2$ (while the gap ratio is still approximately given by RMT value $\overline r\approx 0.531$, see Fig.~\ref{fig:mag_hist_2}(a)). For larger $W$, the probability distribution $\mathcal{P}(s^z)$ starts to resemble the expected Gaussian curve (e.g., at $W=2.1$ and $2.3$). The $\mathcal{P}(s^z)$ distribution starts to admit the U-shape characteristic for the MBL regime only at much larger amplitudes of the QP potential, for instance $W\approx 3.6$ for the system-size $L=18$.

Moreover, the probability distribution 
$\mathcal{P}(|s^z_{l} - s^z_{l+1}|)$ for the nearest-neighbor spin-differences $|s^z_{l} - s^z_{l+1}|$ also shows a peak at non-zero $|s^z_{l} - s^z_{l+1}|$ deep in the ergodic regime where $\mathcal{P}(s^z)$ shows the characteristic double-peaked structure (compare Fig.~\ref{fig:mag_hist_2}(b)). This shows that the nearest-neighbor spins are more likely to be in opposite orientations even in the ergodic phase of Heisenberg model with QP potential.
This behavior is particularly exotic since the ordering of adjacent spins arises for states in the middle of the spectrum of a system with average gap ratio $\overline r$ in agreement with RMT.

The distribution of $\mathcal{P}(s^z)$ is qualitatively similar to a distribution of field differences $h_l-h_{l+1}\equiv \varepsilon $ at neighboring sites, which for the QP potential \eqref{qppot} is given by $P(\varepsilon)=(2 \pi \sin(\pi k ) )^{-1}  (1-(\varepsilon / 2 \pi \sin(\pi k )^2)^{-1/2}$. The distribution $P(\varepsilon)$ has characteristic peaks at $\epsilon = \pm 2 \pi \sin(\pi k )$. The value of $\varepsilon$  determines how resonant the local tunneling is, strongly affecting system properties both for non-interacting models \cite{Guarrera07} as well as for the MBL transition \cite{Doggen19}. It seems reasonable to us that the characteristic shape of $P(\varepsilon)$ for the QP potential determines the double peak structure of $\mathcal{P}(s^z)$. However, we are not able to pin-point a precise mechanism leading to the characteristic shape of $\mathcal{P}(s^z)$ deep in the ergodic regime of the QP system.

The conclusions relevant for the finite-size scaling analysis in the rest of this work are: (i) the behavior of $\mathcal{P}(s^z)$ in QP system is much different than in the random case preventing us from analyzing the ergodic-MBL transition with the chain breaking mechanism of \cite{Laflorencie20}; (ii) the characteristic for the QP potential structure of $\mathcal{P}(s^z)$ appears only if the system is sufficiently large, hence we consider mostly $L\geq16$ in our finite size scaling analysis.  This choice is also motivated by the fact that the average gap-ratio $\bar{r}$ does not vary smoothly with $W$ for $L \leq 14$.

\section{Finite-size scaling analysis}
\label{sec:scaling}

We now  present  the  results of the finite-size scaling of the average gap-ratio and the half-chain EE in the QP Heisenberg chain. First, we analyze the system by considering a fixed critical point $W^*$ independent of the system size. Then, we consider more general scenarios with different size-dependent functional forms of the critical disorder strength.

The basic principle for finite-size scaling is that near the critical point $W^*$ the correlation length $\xi$ diverges 
\cite{Cardy96} either according to \ref{eq:power_law} for a standard second-order phase transition or according to \ref{eq:BKT} for a BKT transition. 
As a result, the normalized observable, $X$, takes a functional form 
\begin{equation}
X=\mathcal{G}(L/\xi),
\end{equation}
where $\mathcal{G}(.)$ is a continuous function.

For the finite-size scaling analysis, we follow the recently proposed method of minimizing a cost-function introduced by Šuntajs et. al. \cite{Suntajs20}.
Unlike other scaling methods it does not assume the observables to take a particular functional form of $L/\xi$ but rather that the observables are simply monotonic functions of $L/\xi$. The cost-function for a quantity $X=\{X_j\}$ that consists of $N$ values at different $W$ and $L$ is defined as
\begin{equation}
\mathcal{C}_X = \frac{\sum^{N-1}_{j=1}\vert X_{j+1}-X_j\vert}{\max\lbrace X_j\rbrace-\min\lbrace X_j\rbrace}-1,
\label{eq:cost}
\end{equation}
where $X_j$'s are sorted according to non-decreasing values of $sgn[W-W^*]L/\xi$. For an ideal collapse with $X$ being a monotonic function of $sgn[W-W^*]L/\xi$, we must have  
$\sum_j \vert X_{j+1}-X_j\vert=\max\lbrace X_j\rbrace-\min\lbrace X_j\rbrace$,
and thus  $\mathcal{C}_X=0$. However, in our case it will suffice that the best collapse corresponds to the global minima of $\mathcal{C}_X$ for different correlation lengths and functional forms of $W^*(L)$. In Appendix~\ref{app:minimize}, we provide details about the numerical optimization of the cost-function performed in this work.

\subsection{Comparisons of finite-size scaling for system-size independent critical disorder strength}

\begin{figure}
\includegraphics[width=\linewidth]{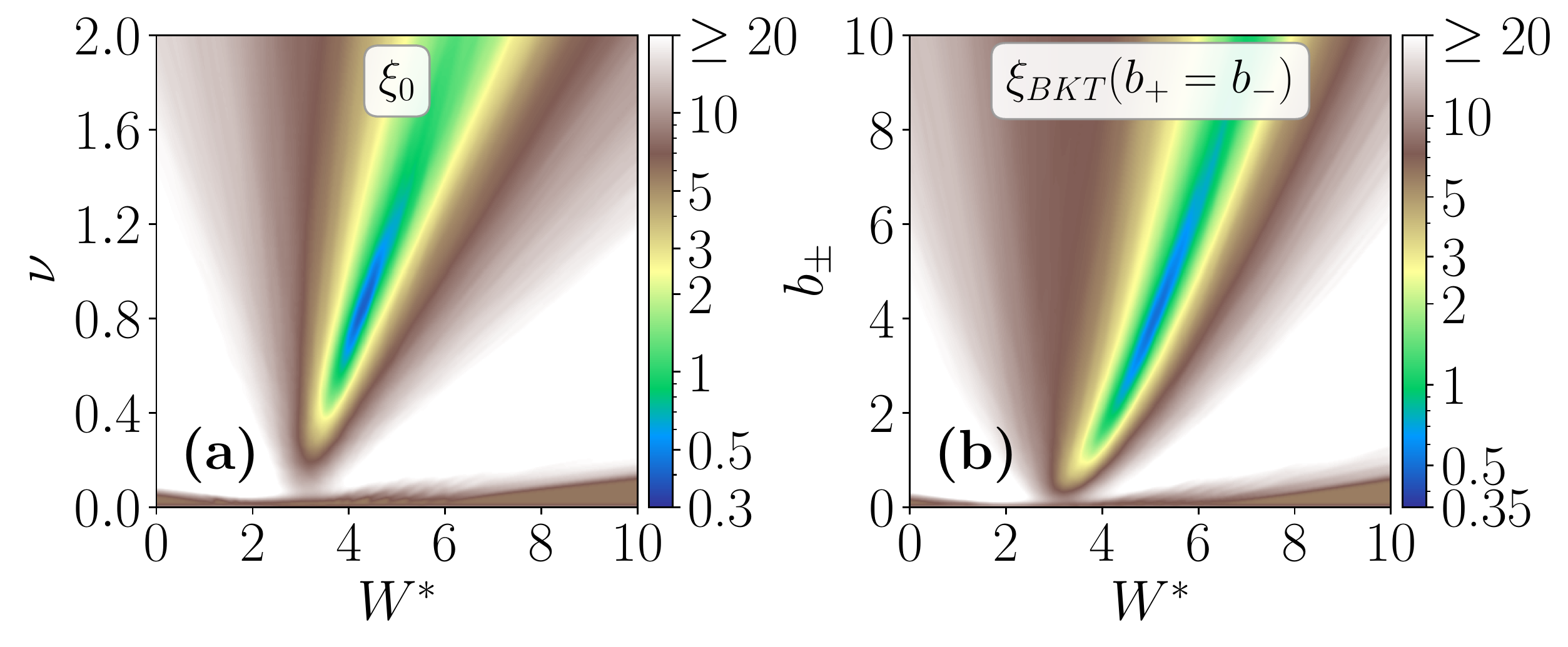}
\caption{{The pattern of the cost-function $\mathcal C_{\mathcal S}$ for the rescaled EE $\mathcal{S}$ as a function of the minimizing parameters $W^*$ and \textbf{(a)} $\nu$ for the correlation length $\xi_0$ or \textbf{(b)} $b_{\pm} = b_+ = b_-$ for the correlation length $\xi_{BKT}$. 
}}
\label{fig:contour}
\end{figure}

Let us first consider the scenario, when the critical disorder strength $W^*$ is independent of the system-size $L$.
We make a  comparison between different types of finite-size scaling
where the correlation lengths are given by power-law ($\xi_0$) and that of a BKT transition ($\xi_{BKT}$). For the BKT scaling, we consider both the constrained symmetric condition $b_+=b_-$ and the free asymmetric condition $b_+\neq b_-$. 
{For a visualization of the cost-function for the rescaled EE, ${\mathcal S}$, we plot    $\mathcal C_{\mathcal S}$
in both cases of the power-law correlation and the BKT one with $b_+=b_-$ in Fig.~\ref{fig:contour}. In  this scenario, definite minima of the cost function in the landscapes of the minimizing parameters can be seen.}
The resulting residual values of the cost-function $\mathcal C_{r}$  ($\mathcal C_{\mathcal S}$) obtained in the minimization procedure for the considered scenarios of the transition and for the average gap-ratio (half-chain EE)
are shown in Table \ref{tab:table1}. 
The finite-size scaling with power-law correlation length $\xi_0$ provides better data collapse  
in comparison to that with $\xi_{BKT}$ as seen by the values of the cost-function.

\begin{table}[tbh]
\caption{\label{tab:table1}%
Cost function $\mathcal{C}_X$ comparison for finite-size scaling with fixed critical disorder strength.
}
\begin{ruledtabular}
\begin{tabular}{lcdr}
\textrm{}&
\textrm{$\xi_0$}&
\multicolumn{1}{c}{\textrm{$\xi_{BKT}(b_+=b_-)$}}&
\textrm{$\xi_{BKT}(b_+\neq b_-)$}\\
\colrule
$\mathcal{C}_r$ & \textbf{0.635} & 0.808 & 0.769 \  \  \   \ \\
$\mathcal{C}_\mathcal{S}$ & \textbf{0.326} & 0.419 & 0.402 \  \  \  \ \\
\end{tabular}
\end{ruledtabular}
\end{table}

\begin{figure}[tbh]
\includegraphics[width=\linewidth]{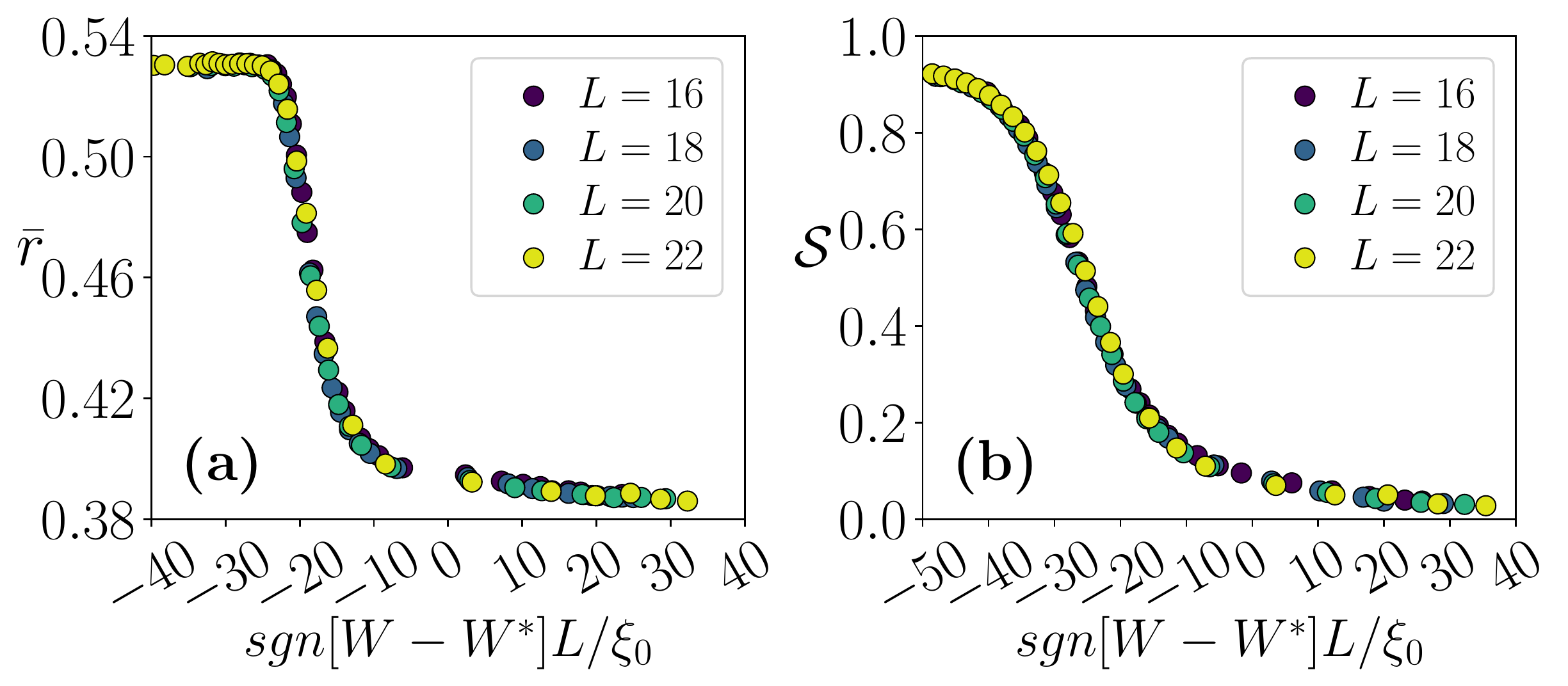}
\caption{ The finite-size scaling for fixed critical disorder strength $W^*$ assuming the power-law divergence of the correlation length $\xi_0$ for (a) the average gap-ratio $\bar{r}$ and (b) the half-chain EE $\mathcal{S}$.
For $\bar{r}$ we find $W^*=3.97$ and $\nu=0.54$ and for $\mathcal{S}$ we get $W^*=4.27$ and $\nu=0.87$.}
\label{best}
\end{figure}

The best data collapses, i.e., with the power-law correlation length $\xi_0$, for each observables are presented in Fig. \ref{best}. The critical disorder strength turns out to be $W^*_{(\bar{r})} = 3.97$ and $W^*_{(\mathcal{S})} = 4.27$ respectively for the average gap-ratio $\bar{r}$ and the half-chain EE $\mathcal{S}$. However, the critical exponent $\nu$ that we extract from the data collapse is 0.54 and 0.87 for $\bar{r}$ and $\mathcal{S}$ respectively. While $\nu_{(\mathcal{S})} \sim 1$ matches pretty well with the recent results for a similar QP chain (with added next-neighbor tunnelings) \cite{Khemani17}, the exponent for $\bar{r}$ ($\nu_{(\bar{r})} \sim 0.5$) contradicts those results. 
Therefore, while $\nu_{(\mathcal{S})}$ is close to obey the Harris-Luck criteria of $\nu > 1$ \cite{Luck93}, $\nu_{(\bar{r})}$ strongly violates it.

For second-order phase transitions the finite size scalings are usually best represented by a fixed critical disorder strength, while for BKT scaling even for system sizes of $\mathcal{O}(1000)$ we expect a drift of the form $\sim 1/\ln(L)^2$ \cite{Bramwell94}. The phenomenological  renormalization group studies find more exotic finite-size scaling for MBL in RD systems \cite{Goremykina19,Dumitrescu19,Morningstar19,Morningstar20}, and the proper mechanism for MBL transition in QP systems are yet to be put forth. Thus, we assume $W^*$ to take different functional forms with the system-size $L$, and continue our analysis on the finite-size scaling assuming $W^*$ now to be dependent on the system-size $L$ in the following subsection.
This is also motivated by similar studies on quenched disordered systems that found a linear drift in critical disorder strength with increasing length \cite{Suntajs20,Sierant20p,Suntajs20e}. While few previous studies have commented on this behavior being relatively weak for QP systems (thus emphasizing that QP systems are relatively stable) \cite{Khemani17, Doggen19, Lee17}, to the best of our knowledge these have not been based on substantial quantitative arguments.

\subsection{Finite-size scaling for system-size dependent critical disorder strength}

We now move to the finite-size scaling analysis of the relevant quantities where we assume that the critical disorder strength has drifts with the system-size $L$. Specifically, we consider the following functional forms for the critical point $W^*$ (we refer to the Appendix~\ref{app:func} for a comparison with alternative system-size dependencies {that asymptotically lead to fixed critical points in the thermodynamic limit}):
\begin{enumerate}
\item a linear drift with respect to $L$,  $W^*=W^0 +W^1L$, as in for uniformly distributed random Heisenberg chain considered in \cite{Suntajs20,Suntajs20e},
\item a logarithmic drift $W^*=W^0+W^1\ln L$,
\item and the most generic $W^* = W^*(L)$ where $W^*(L)$ are chosen individually for each $L$.
\end{enumerate}
As before, we consider both power-law \ref{eq:power_law} and (both symmetric and asymmetric) BKT \ref{eq:BKT} scenarios for the ergodic-MBL transition.

\begin{table}[htb]
\caption{\label{tab:table2}%
Cost-function $\mathcal{C}_X$ comparison for finite-size scaling with system-size dependent drifts in the critical disorder strength.
}
\begin{ruledtabular}
\begin{tabular}{lccr}
{ } &
\textrm{$W^0+W^1 L$}&
\multicolumn{1}{c}{\textrm{$W^0+W^1 \ln L$}}&
\textrm{$W^*(L)$}\\
\colrule
$\mathcal{C}_r[\xi_0]$ & 0.200 & 0.192 & 0.192 \\
$\mathcal{C}_r[\xi_{BKT}(b_+=b_-)]$ & 0.221 & \textbf{0.147} & \textbf{0.144} \\
$\mathcal{C}_r[\xi_{BKT}(b_+\neq b_-)]$ & 0.206 & \textbf{0.143} & \textbf{0.139}\\
\hline
$\mathcal{C}_\mathcal{S}[\xi_{0}]$ & 0.076 & 0.076 & 0.076 \\
$\mathcal{C}_\mathcal{S}[\xi_{BKT}(b_+=b_-)]$ & 0.151 & \textbf{0.039} & \textbf{0.035} \\
$\mathcal{C}_\mathcal{S}[\xi_{BKT}(b_+\neq b_-)]$ & 0.117 & \textbf{0.034} & \textbf{0.026}\\
\end{tabular}
\end{ruledtabular}
\end{table}

\begin{figure}%
\centering
\includegraphics[width=\linewidth]{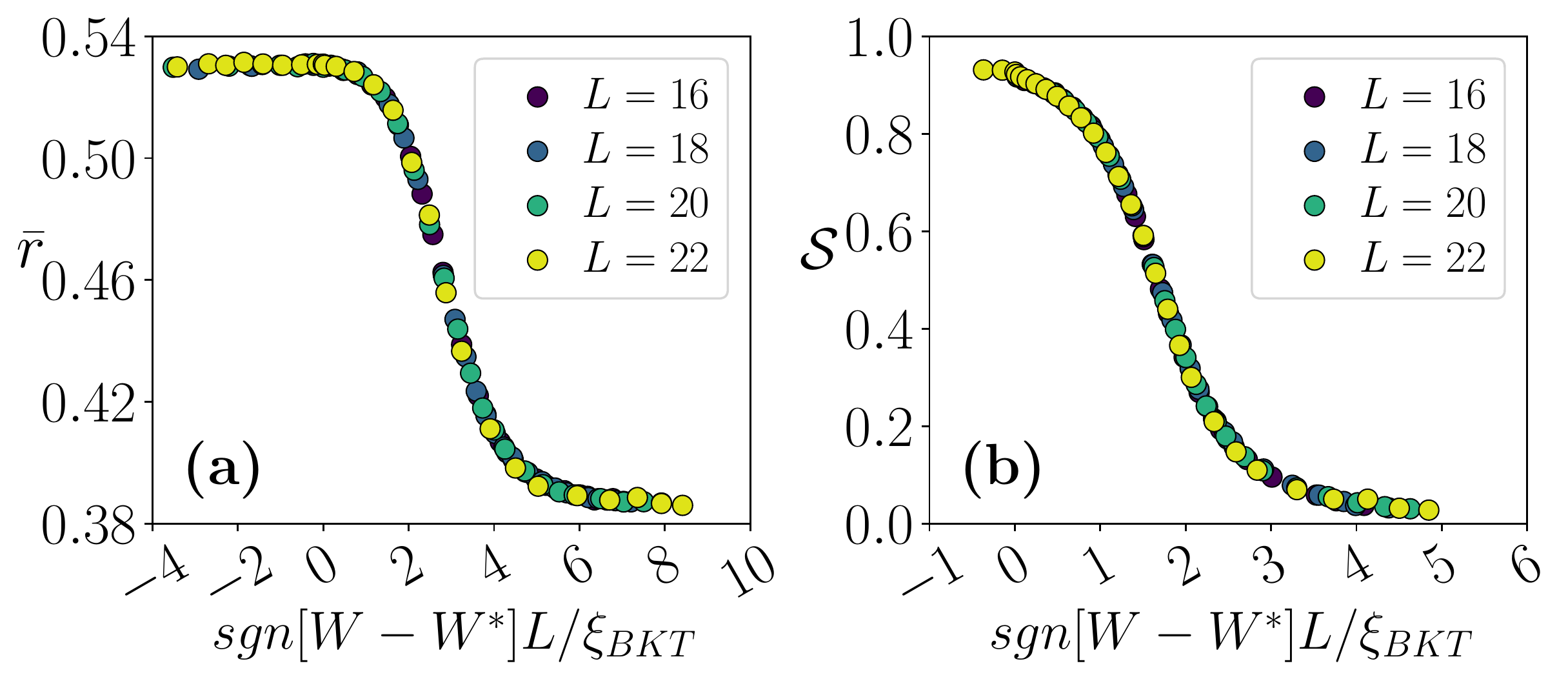}
\caption{
Finite-size scaling with the logarithmic drift in the critical disorder strength having (symmetric) BKT correlation length  $\xi_{BKT}(b_-=b_+)$ for (a) $\bar{r}$ and (b) $\mathcal{S}$.
For $\bar{r}$ we find $W^0 = -6.05$, $W^1=2.77$ and $b_{\pm}=1.79$ while for $\mathcal{S}$ we get $W^0=-8.19$, $W^1=3.17$ and $b_{\pm}=3.17$.}
\label{bestwithL}
\end{figure}

The corresponding values for the cost-function for the mentioned scenarios are presented in Table~\ref{tab:table2} for comparison. Importantly, by comparing the values of the cost-function with those presented in Table~\ref{tab:table1}, we find that the finite-size scaling with system-size dependent critical disorder strength produces much better result than that with fixed critical point.
We note that the scaling with power-law type correlation length still works better for the linear drift in critical point ($W^*=W^0 +W^1L$) compared to the BKT ones. Interestingly, the performance of scaling with the linear drift is significantly worse than the scalings with the logarithmic drift ($W^*=W^0+W^1\ln L$) and for generic system-size dependence $W^* = W^*(L)$. This is especially relevant for the BKT scenario of the transition (compare the data in Table~\ref{tab:table2}).
These observations unravel significant differences between the ergodic-MBL transitions in the Heisenberg model with QP potential and in the model with RD where the scaling with a linear drift having BKT type divergence in correlation length performs the best \cite{Suntajs20}.

In Fig. \ref{bestwithL}, we show the finite-size scaling with the correlation length
$\xi_{BKT}(b_-=b_+)$ with the logarithmic drift for $\bar{r}$ and $\mathcal{S}$. However, for symmetric BKT scaling ($b_- = b_+$),
we find that the values of $(W^0, W^1)$, as presented in the caption of Fig.~\ref{bestwithL}, are not similar for $\bar{r}$ and $\mathcal{S}$. 
On the other hand, while the condition $b_-\neq b_+$ only results in marginal improvements in the cost-function (see Table~\ref{tab:table2}),
the extracted values of $(W^0,W^1)$ are similar for $\bar{r} \rightarrow (-6.57,2.92)$ and for $\mathcal{S} \rightarrow (-6.28,2.70)$. 
The non-universal constants $(b_-,b_+)$ obtained with the logarithmic drift are found to be $(23.72,1.82)_{(\bar{r})}$ and $(0.14,2.75)_{(\mathcal{S})}$ respectively for $\bar{r}$ and $\mathcal{S}$ (similar values are also obtained for generic $W^*=W^*(L)$). 
Note here that while the values of $b_+$, for both $\bar{r}$ and $\mathcal{S}$, are quite similar to those of the constant $b_{\pm}$ extracted for the symmetric BKT scaling, the values of $b_-$ are markedly different which seems to be unusual. Nevertheless, these observations suggests that the finite-size scaling in two sides of the critical point is very much asymmetrical. Such an asymmetric scaling at the ergodic-MBL transition has also been hinted for the Heisenberg system with RD \cite{Laflorencie20} (see also \cite{Mace19b}).

However, the minimization of the cost-function for the BKT scaling with the asymmetric free condition $b_-\neq b_+$ relies on a number of technical subtleties, and the corresponding results must be taken with caution. Specifically, we find that there are multitudes of degeneracies in the values of the minimized cost-function that appear for different sets of the values of the parameters $(W^0, W^1, b_-, b_+)$. The numbers quoted for the BKT scaling with the free condition $(b_-\neq b_+)$ in the preceding paragraph are just one example of $(W^0, W^1, b_-, b_+)$ that minimize $\mathcal{C}_X[\xi_{BKT}(b_- \neq b_+)]$. However, we also notice that $(W^0, W^1, b_+)$ remains similar in values between different minimizing sets, but $b_-$ fluctuates a lot in values between these sets. For these reasons, we now continue the finite-size scaling having BKT correlation length $\xi_{BKT}$
with the data taken only from the localized side, i.e., only with the constant $b_+$. Such analysis is also motivated by the results of \cite{Laflorencie20} that shows that the BKT type finite-size scaling performs best when looking from the localized side in random disordered Heisenberg chain, {while a volumic scaling is revealed in the ergodic side where the scaling variable is a ratio between the corresponding Hilbert space dimension and a disorder-dependent non-ergodicity volume that diverges exponentially at the critical point}.

\begin{table}[htb]
\caption{\label{tab:table_loc}%
Cost-function $\mathcal{C}_X$ for the finite-size scaling performed on the localized side.
}
\begin{ruledtabular}
\begin{tabular}{lccr}
{ } &
\textrm{$W^0+W^1 L$}&
\multicolumn{1}{c}{\textrm{$W^0+W^1 \ln L$}}&
\textrm{$W^*(L)$}\\
\colrule
$\mathcal{C}_r[\xi_{BKT}]$ & 0.074 & 0.07 & 0.07 \\
$\mathcal{C}_\mathcal{S}[\xi_{BKT}]$ & 0.066 & 0.033 & 0.026 \\
$\mathcal{C}_\mathcal{S}[\xi_{BKT}]_{\{L \in [12, 22]\}}$ & 0.4 & 0.26 & 0.25 \\
\end{tabular}
\end{ruledtabular}
\end{table}

\begin{figure}[tbh]
\includegraphics[width=\linewidth]{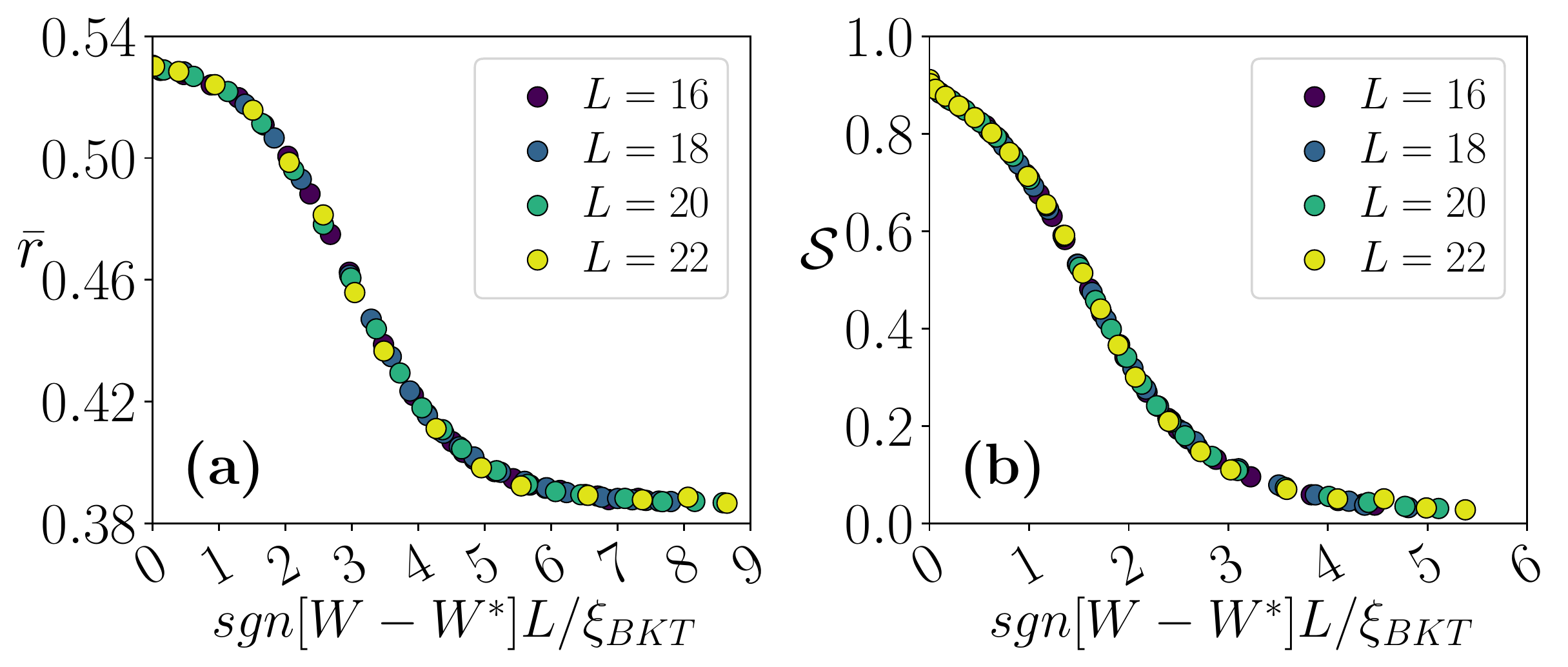}
\caption{
Finite-size scaling in the localized regime with the logarithmic drift in the critical disorder strength having BKT correlation length  $\xi_{BKT}(b_+)$ for (a) $\bar{r}$ and (b) $\mathcal{S}$.
For $\bar{r}$ we find $W^0 = -5.59$, $W^1= 2.66$ and $b_{+}=1.61$ while for $\mathcal{S}$ we get $W^0=-6.62$, $W^1=2.78$ and $b_{+}=2.83$.}
\label{bestwithL_loc}
\end{figure}

\begin{figure}[tbh]
\centering
\includegraphics[width=\linewidth]{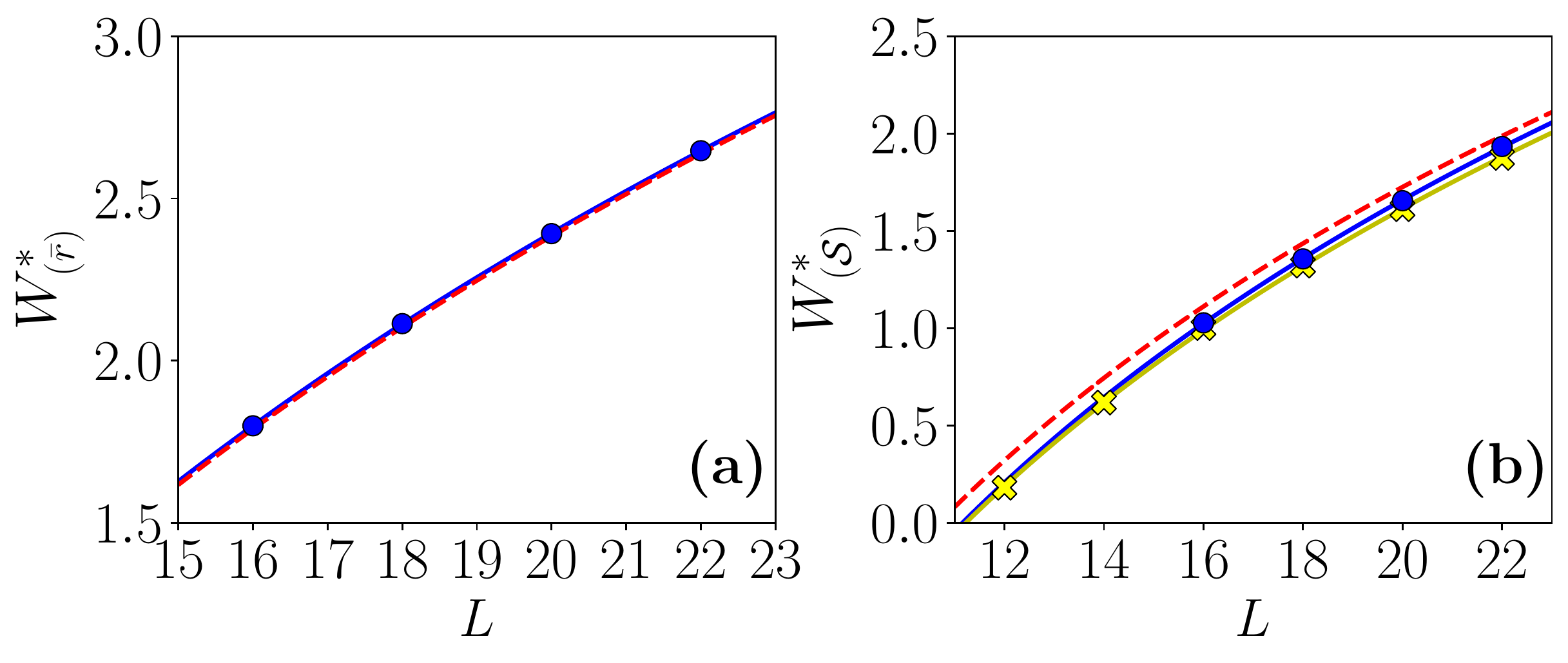}
\caption{
The critical points $W^*$ obtained for different values of $L$ (shown by the solid blue circles) extracted from the finite-size scaling in the localized regime having BKT correlation length $\xi_{BKT}(b_+)$ assuming generic system-size dependence $W^*=W^*(L)$ for (a) average gap-ratio $\bar{r}$ and (b) half-chain EE $\mathcal{S}$.
{In (b), we also perform the same analysis including the data for $L=12$ and $14$, and the extracted values of $W^*(L)$ are presented by the yellow crosses.
Solid lines represent the fit of the $W^*$ to the functional form $W^0 + W^1\ln L$ (see the text), and the red dashed lines correspond to the values obtained by minimizing the cost-function for the logarithmic drift $W^* = W^0 + W^1\ln L$ (see Fig.~\ref{bestwithL_loc}).}
}
\label{gen}
\end{figure}

For the finite-size scaling only from the localized side, we keep the definition of the cost-function (Eq.~\eqref{eq:cost}) unchanged but only include the data points $\{X_j\}$, $X \in \{\bar{r}, \mathcal{S}\}$, that satisfy $sgn[W-W^*]L/\xi > 0$. In Table~\ref{tab:table_loc} we display the corresponding values of the cost-function for system-size dependent critical disorder strengths as in Table~\ref{tab:table2}. 
As before, we again find that the logarithmic drift as well as generic $W^*=W^*(L)$ outperforms the scaling with the linear drift.
Figure~\ref{bestwithL_loc} shows the data collapse for the finite-size scaling with the logarithmic drift for both $\bar{r}$ and $\mathcal{S}$. Corresponding minimizing parameters $(W^0, W^1, b_+)$ are given in the figure caption.
In Fig. \ref{gen} we show the different $W^*$ values obtained by minimizing the cost-function for generic $W^*=W^*(L)$ for finite-size scaling on the localized side with BKT correlation length. Moreover, we find that these values fit perfectly with a logarithmic functional form, again indicating a logarithmic drift in the given system-sizes. The fit parameters obtained are also very similar to the the ones obtained by minimizing with the logarithmic drift, especially for $\bar{r}$ where both are effectively identical.
To understand better the drift of the critical point with the system size, we perform the same analysis with the inclusion of the data for $L=12$ and $14$ for the half-chain EE $\mathcal{S}$ (see Fig.~\ref{gen}(b)). In this case also, we find that the scaling is best described by the critical point having a logarithmic drift or the generic $W^* = W^*(L)$ (see Table~\ref{tab:table_loc}).
In fact, both scenarios 
perfectly fit to the functional form $W^0 + W^1\ln L$  as shown by the bold lines in Fig.~\ref{gen}. 
For $\bar{r}$ we find $W^0=-5.58$ and $W^1=2.66$, while for $\mathcal{S}$ we get $W^0=-6.86$ and $W^1=2.84$ from the fit.
In (b), we also perform the same analysis including the data for $L=12$ and $14$, and the extracted values of $W^*(L)$ are presented by the yellow crosses. In this also, $W^*(L)$ fits to logarithmic functional form with $W^0 = -6.78$ and $W^1 = 2.8$ which are very close in values to the previous case.

\begin{figure}%
	\centering
	\includegraphics[width=\linewidth]{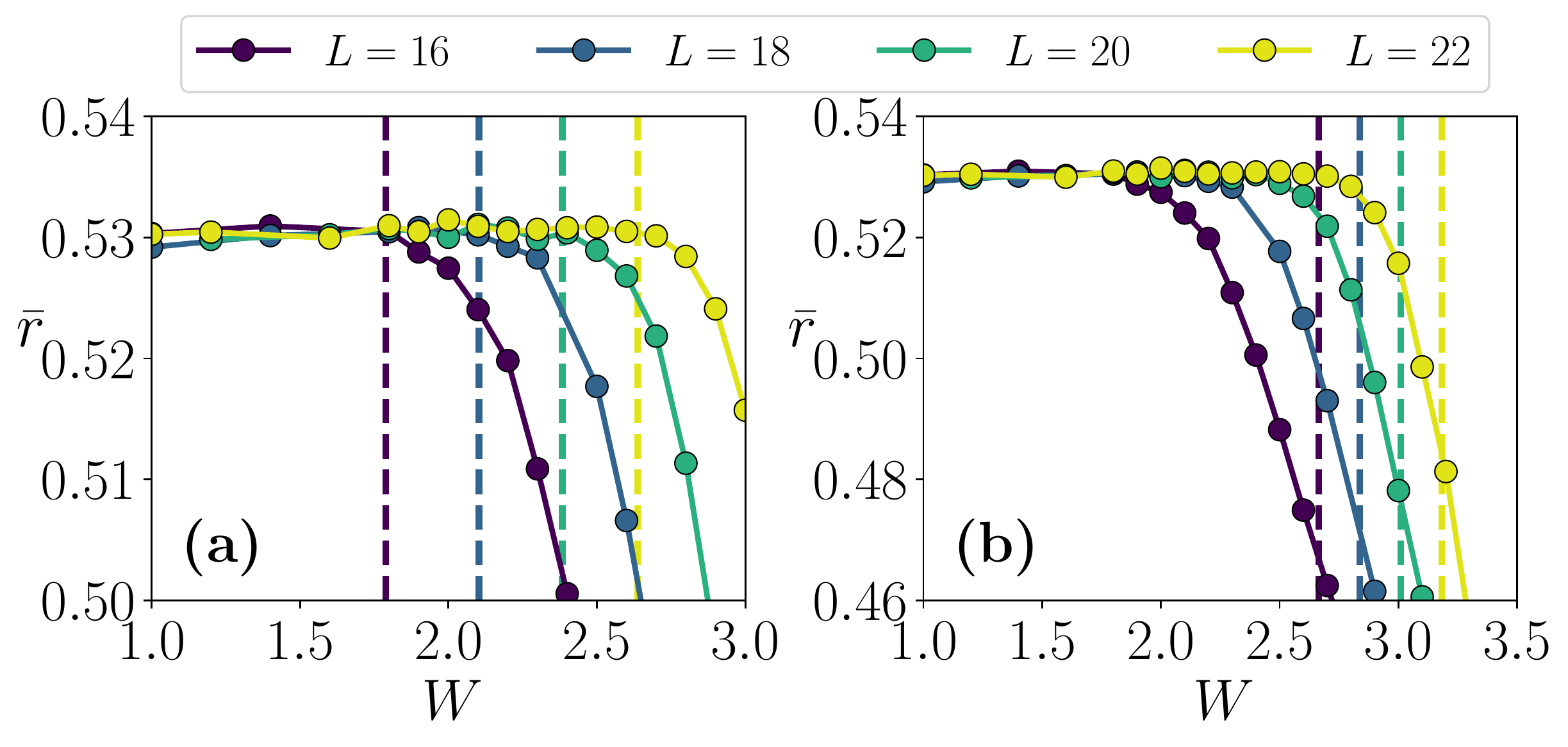} \\
	\includegraphics[width=0.5\linewidth]{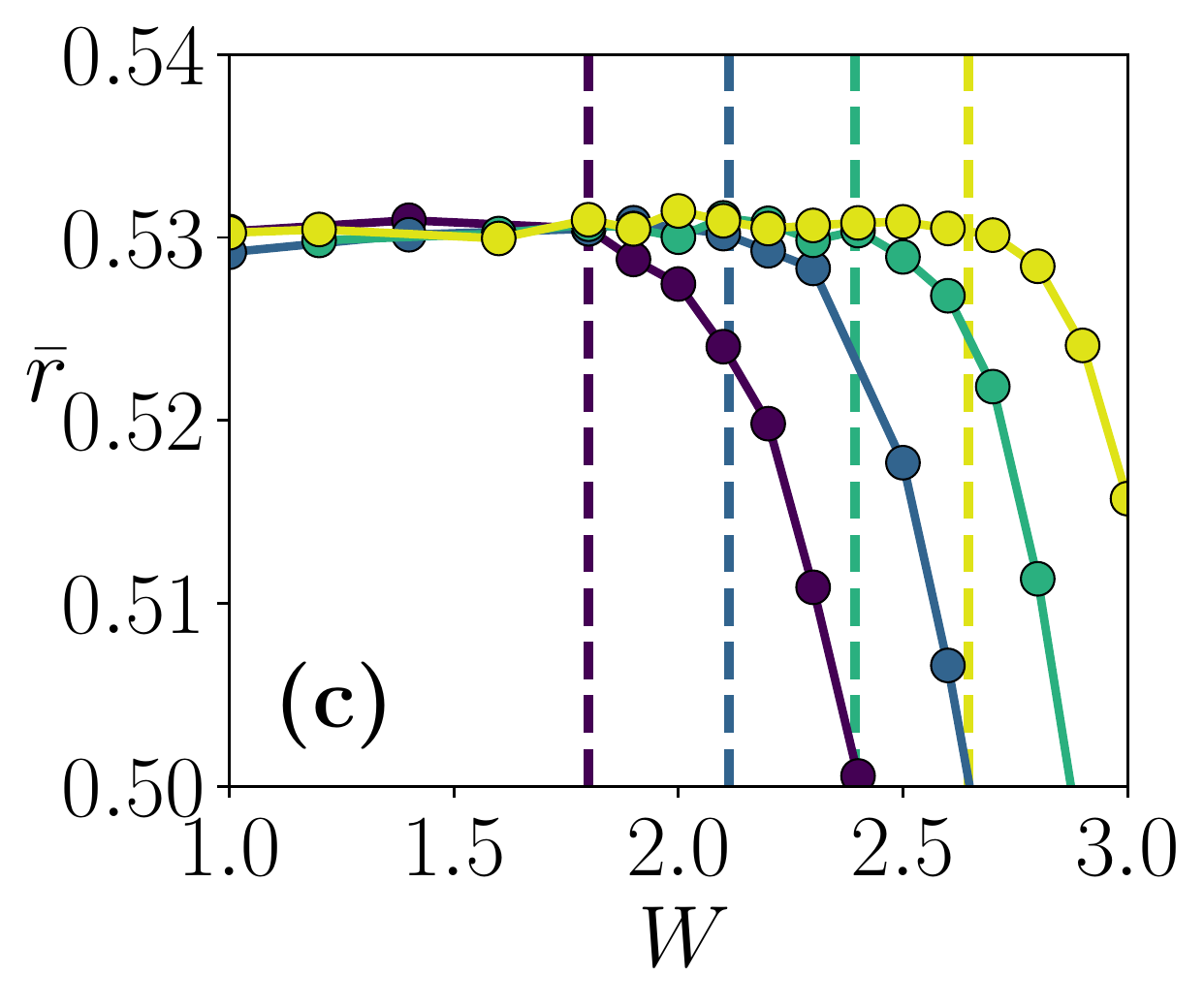}
	\caption{The critical disorder strengths for different system-sizes, as shown by the vertical dashed lines, obtained from the BKT scaling of  $\bar{r}$ in the localized regime by assuming (a) a logarithmic drift and (b) a linear drift of the critical point with respect to the system-size, as well as by considering (c) generic $W^*=W^*(L)$.}
	\label{cuts}
\end{figure}

To further illustrate the system size behavior of the critical point we compare, in Fig.~\ref{cuts},  the critical disorder strengths for the BKT type scaling of  $\bar{r}$ in the localized regime considering different functional forms of the critical point. Interestingly, we get that for the logarithmic drift and for the generic $W^*=W^*(L)$ the critical points are located at values for which the average gap-ratio $\bar{r}$ just starts to deviate from the RMT value $(\sim 0.531)$.

\subsection{Summary of the finite-size scaling analysis}

Our analysis shows that when the critical disorder strength is considered to be system-size independent, the power-law scaling provides better results compared to the BKT scaling. However, the critical exponents that we obtain from the scaling of the average gap-ratio $\bar{r}$ and half-chain entanglement entropy $\mathcal{S}$ are found to be $\nu_{(\bar{r})} \sim 0.5$ and $\nu_{(\mathcal{S})} \sim 1$ respectively. While the latter is roughly consistent with the Harris-Luck criterion for the QP systems \cite{Luck93}, the former clearly breaks the criterion. This suggests that the transition in this QP system is not stable within the system-sizes considered here, in contradiction with the predictions of \cite{Khemani17, Lee17}.

On the other hand, by assuming that 
the critical point drifts with increasing system-size, we find that the finite-size scaling performs best for the BKT scenario with a logarithmic drift (within available system sizes) of the critical disorder strength. 
The system-size dependent critical disorder strengths that we extract from the scaling of $\bar{r}$ and $\mathcal{S}$ are respectively $W^*_{(\bar{r})} = -5.58 + 2.66 \ln L$
 and $W^*_{(\mathcal{S})} = -6.86 +  2.84 \ln L$  (see Fig.~\ref{gen} {and the corresponding discussions in the text}), that results into $W^*_{(\bar{r})} \simeq 2.64$ and $W^*_{(\mathcal{S})} \simeq 1.97$ for a system of size $L=22$, both being very close to the ergodic regime.
On the other hand, since our study is performed for a very narrow window in the system-sizes (specifically, $L \in [16, 22]$), we cannot be sure that the drift is indeed logarithmic. By analyzing the data of the entanglement entropy in the range $L \in [12, 22]$, we see that the logarithmic drift still persists in this wider range.
This shows clearly that the critical disorder strength is certainly \textit{sub-linear} with respect to the system-size in contrast to the situation for the Heisenberg chain with RD. 
{ 
Interestingly, a previous study done on the same model has predicted $W^* \simeq 4.8 \pm 0.5$  using time-evolution of large systems ($L=50$)
\cite{Doggen19} that closely matches our prediction ($W^*_{(\bar{r})} \simeq 4.83$ and $W^*_{(\mathcal{S})} \simeq 4.25$ for $L=50$)
when the logarithmic drift of $W^*$ is extrapolated to $L=50$. This is also consistent with the value of critical disorder strength reported recently in \cite{Singh21}.}

\section{Concluding remarks}
\label{sec:conclu}

In this work, we have performed a detailed analysis of the ergodic-MBL transition in finite-size quasi-periodic Heisenberg model. In the ergodic regime, instead of the expected Gaussian probability distribution of on-site magnetization $\mathcal P(s_z)$,
we found an exotic double-peak structure of $\mathcal P(s_z)$. This highlight the importance of correlations in the QP potential showing differences between QP and RD models even in the ergodic regime.

A detailed analysis of the power-law \ref{eq:power_law} and {the} BKT \ref{eq:BKT} scenarios suggests that the  latter is better suited to describe the system size scaling at the transition to MBL phase. The similar approach points towards the BKT scaling at the transition in RD systems \cite{Suntajs20}. Taken literally, this suggests that MBL transitions belong to the same BKT universality class, in contradiction with \cite{Khemani17}. However, this result must be interpreted with caution  due to the narrow interval of system sizes available both for QP and RD systems.

Our analysis of system size drifts of the critical disorder strength $W^*$ in the QP model shows that $W^*$ increases sub-linearly with $L$ and that the finite size effects are less severe than in the RD case. The increased stability of MBL regime of QP systems in comparison to RD case is further supported by results for time-evolution  \cite{sierzak}. This suggests that QP systems may be better suited to understand the asymptotic behavior of the MBL phase~\cite{Morningstar21}.

\begin{acknowledgments}
The numerical calculations have been possible thanks to PL-Grid Infrastructure.
This research has been supported  by
 National Science Centre (Poland) under project 2019/35/B/ST2/00034 (A.S.A., J.Z.). The work of T.C. was realised within the QuantERA grant QTFLAG, financed by  National Science Centre (Poland)  via grant 2017/25/Z/ST2/03029.  P.S. acknowledges the support of  Foundation  for
Polish   Science   (FNP)   through   scholarship   START.
\end{acknowledgments}

\appendix

\section{Minimization of the cost-function }
\label{app:minimize}

For minimizing the cost-function, we employ a combination of the differential evolution method and the Nelder-Mead simplex algorithm implemented in SciPy \cite{Scipy} respectively for global and local optimizations.
For each minimization procedure, we run a series of independent differential evolution algorithms each having different random seed followed by  a Nelder-Mead simplex method. We employ $\sim 100$ such independent realizations to find the optimal solution. At  each 
differential evolution method we allow upto $10^4$ iterations with a relative tolerance of convergence $10^{-4}$ and with a population size of $10^3$.

{
However, one crucial point in our analysis is about the stability of results obtained
{in} the {minimization of the} cost-function keeping in mind that the data $X_i$ used (either $\bar r(L,W)$ or the rescaled entropies) are obtained with some statistical error  due to disorder averaging. To test that we consider the rescaled EE example and assume that {$\mathcal{S}$} values for each $L$ and $W$ are Gaussian distributed with centers at the obtained mean values and the standard deviations given by errors of these means. Probing the data over 1000 such {Gaussian distributed} realizations and performing the cost-function analysis, we obtain for the symmetric BKT case, {i.e.,} the one depicted in
Fig.~\ref{bestwithL}\textbf{(b)}, the value $\mathcal{C}_{\mathcal{S}}[\xi_{BKT}(b_+=b_-)] = 0.040 \pm 0.004$ which matches 0.039 value in Table~\ref{tab:table2}. Similarly, taking localized regime only, {i.e.,} corresponding to  Fig.~\ref{bestwithL_loc}\textbf{(b)}, we obtain
$\mathcal{C}_{\mathcal{S}}[\xi_{BKT}] = 0.034 \pm 0.003$ as compared to the corresponding value in Table~\ref{tab:table_loc} of 0.033.
{Moreover, in these case, the minimizing parameters remains virtually unchanged upto two decimal places.
These agreements in values as well as such small errors in the cost-function that occur in the third decimal place confirms that the cost-function analysis is very stable with respect to the statistical errors that come with disorder averaging.}
}

\section{Asymptotic functional forms for critical disorder strength}
\label{app:func}

\begin{table}
\caption{\label{tab:table4}%
Cost function $\mathcal{C}_X$ comparison for finite-size scaling with asymptotic functional forms of critical disorder strength.
}
\begin{ruledtabular}
\begin{tabular}{lccr}
\textrm{}&
\textrm{$\xi_0$}&
\multicolumn{1}{c}{\textrm{$\xi_{BKT}$}}\\
\colrule
$\mathcal{C}_{\bar{r}}$: $W_0$ &0.635  & 0.808  \\
$\mathcal{C}_{\bar{r}}$: $W_0+W_1/L$ & 0.633 & 0.788  \\
$\mathcal{C}_{\bar{r}}$: $W_0+W_1/(\ln(L))$ & 0.554 &0.737  \\
\hline
$\mathcal{C}_\mathcal{S}$: $W_0$ & 0.326 & 0.419 \\
$\mathcal{C}_\mathcal{S}$: $W_0+W_1/L$ & 0.307 &  0.402\\
$\mathcal{C}_\mathcal{S}$: $W_0+W_1/(\ln(L))$ &  0.286 & 0.419 \\
\end{tabular}
\end{ruledtabular}
\end{table}

On making a cost-function analysis with functional forms that asymptotically leads to a finite value, we find the cost-function improvements over the fixed critical point to be marginal in comparison with linear or logarithmic drift. This is shown in TABLE \ref{tab:table4}.


%

\end{document}